\documentclass{jfm}
\usepackage{epstopdf, epsfig}
\usepackage{amsmath}
\usepackage{siunitx}
\usepackage{natbib}

\shorttitle{Lift Force on a Moving Intruder in Granular Shear Flow}
\shortauthor{H. He, Q. Zhang, J. M. Ottino, P. B. Umbanhowar, and R. M. Lueptow}

\title{Lift Force on a Moving Intruder in Granular Shear Flow}

\author{Hantao He\aff{1},
  Qiong Zhang\aff{1},
  Julio M. Ottino\aff{1,2,3},
  Paul B. Umbanhowar\aff{1},
 \and Richard M. Lueptow\aff{1,2,3}
 \corresp{\email{r-lueptow@northwestern.edu}}}

\affiliation{\aff{1}Department of Mechanical Engineering, Northwestern University, Evanston, Illinois 60208, USA
\aff{2}Department of Chemical and Biological Engineering, Northwestern University, Evanston, Illinois 60208, USA
\aff{3}Northwestern Institute on Complex Systems (NICO), Northwestern University, Evanston, Illinois 60208, USA}

\begin{document}

\maketitle

\begin{abstract}
Lift and drag forces on moving intruders in granular materials are of fundamental interest. While the drag force on an intruder in granular flow has been studied, the few studies characterizing the lift force explore a relatively limited range of parameters. Here we use discrete element method (DEM) simulations to measure the lift force, $F_\mathrm{L}$, on a spherical intruder in a uniformly sheared bed of smaller spheres for a range of intruder slip velocities, $u_\mathrm{s}$, relative to the unperturbed flow. In what at first appears as a puzzling result, $F_\mathrm{L}$ in granular shear flow acts in the opposite direction to the Saffman lift force on a sphere in a sheared fluid at low $u_\mathrm{s}$, reaches a maximum value, and then decreases, eventually reversing direction and becoming comparable to $F_\mathrm{L}$ for a fluid. This non-monotonic response holds over a range of flow conditions, and the $F_\mathrm{L}$ versus $u_\mathrm{s}$ data can be collapsed by scaling both quantities using the particle sizes, shear rate, and overburden pressure. Analogous fluid simulations demonstrate that the flow field around the intruder particle is similar in the granular and fluid cases. However, the shear stress acting on the intruder in a granular shear flow is much less than that in a fluid shear flow. This difference, combined with a void region behind the intruder in granular flow, which alters the pressure and shear stress on the trailing side of the intruder, significantly changes the lift-force inducing stresses acting on the intruder between the granular and fluid cases. 
\end{abstract}

\section{Introduction}
A fundamental question in both granular and fluid flows relates to the nature and magnitude of forces on a single intruder particle in the flow. In fluid flows, these forces include lift and drag. At the simplest level, the forces on an intruder in a fluid at any instant are body forces, such as weight, and surface forces, characterized by integrating the surface normal (pressure) and shear stresses over the intruder.  Although lift and drag relationships are well-established for intruders in fluid flows, a similar level of understanding is lacking for granular flows. Hence, recent research has focused on the forces on a single intruder particle in dense granular flow. As in a fluid, the combination of body forces and surface forces result in the net force acting on an intruder in a granular flow. However, the challenge lies in isolating those forces when they come from discrete contact interactions with the surrounding particles in a granular flow, unlike interactions with a continuous medium in a fluid flow.

Beyond the fundamental question of what forces act on an intruder particle in granular flows, these forces are also a key part of granular rheology~\citep{nichol2010flow,reddy2011evidence} and are crucial in models of interspecies interactions~\citep{jenkins2002segregation,gray2018particle,duan2020segregation,bancroft2021drag}. They are of practical importance in many applications including particle mixing and segregation~\citep{jenkins2002segregation,tripathi2011numerical, gray2018particle,jing2021unified, jing2022drag}, impact and penetration~\citep{umbanhowar2009enhanced,umbanhowar2010granular}, and animal and robotic locomotion in granular media~\citep{gravish2010force,li2013terradynamics}.

With the advance of numerical methods and computational capabilities for simulating dense granular flows, especially the discrete element method (DEM), which isolates particle-level behavior (particle motions and inter-particle contact forces) in simulations that are challenging or impossible to replicate in actual experiments~\citep{jing2017micromechanical}, progress is being made in understanding the forces on single intruders in granular flows~\citep{ding2011drag,tripathi2011numerical,jing2022drag}. Similar to the case of an intruder in a fluid flow, it is useful to partition the net forces into components, e.g., buoyancy, drag, and lift forces.  For instance, the drag force on a spherical intruder that is larger than the bed particles in a granular shear flow follows a Stokes-like drag relation over five orders of magnitude the intruder Reynolds number~\citep{jing2022drag}. A similar Stokes relation for the drag force describes the situation for heavy intruder particles in size monodisperse chute flows~\citep{tripathi2011numerical} and planar simple shear flow~\citep{liu2017transport}. Similarly, an intruder differing in size from the surrounding flowing bed particles in the presence of gravity experiences a buoyancy force similar to Archimedes' principle but modified by the packing fraction in the granular flow~\citep{huerta2005archimedes, tripathi2011numerical,tripathi2013density,lantman2021granular}. In general, the net segregation force that typically results in large particles rising and small particles sinking in a flowing granular material results from the combined effect of buoyancy as well as shear gradients in the flow~\citep{jing2021unified, guillard2016scaling}.


Compared with drag and segregation forces, significantly less attention has been devoted to the lift force, $F_\mathrm{L}$, in granular flows, which occurs when an intruder moves with a ``slip" velocity, $u_\mathrm{s}$,  relative to the base local streamwise flow velocity.  In one study on this topic, \citet{van2018segregation} measured the lift force on a spherical intruder in a gravity-driven chute flow (non-uniform shear flow) and related it to the intruder's slip velocity using the Saffman lift model, which predicts the lift force on a sphere in a uniformly sheared fluid~\citep{saffman1965lift}.  However, the $u_\mathrm{s}$ values in \citet{van2018segregation} are not directly controlled and are small with large uncertainties. More recently, \citet{yennemadi2023drag} investigated lift forces with imposed streamwise drag forces in a gravity-free granular flow with a uniform shear rate, $\dot\gamma$, and a gravity-driven chute flow.  In contrast to \citet{van2018segregation}, they found that the lift force is inconsistent with the Saffman lift force model, but instead proposed that it is proportional to the collisional stress difference between the upper and lower surfaces of the intruder.  Again, however, the slip velocities in this work are small (i.e.,  $u_\mathrm{s} \ll d_\mathrm{i} \dot\gamma$, where $d_\mathrm{i}$ is the intruder diameter). 

In addition to its fundamental interest and importance, the lift force in granular flows is potentially of practical importance for a more complete understanding of particle segregation in granular flows~\citep{gray2018particle, umbanhowar2019modeling}. In fact, the work of \citet{van2018segregation} frames the non-buoyancy forces driving granular segregation entirely in terms of the slip-driven lift force in contrast to the more typical approach that assumes shear gradients or granular temperature gradients drive segregation, see, e.g., \citet{jing2021unified}.  Furthermore, the lift force in granular flows might also affect the accuracy of measurement of other forces (e.g., drag and buoyancy) in granular flows. For instance, \citet{jing2022drag} noted that their measurement of the drag force might be affected by lift that is induced by the velocity difference perpendicular to the direction the intruder is dragged through the shear flow. Therefore, a systematic and detailed investigation of the lift force in granular flows is also of more practical importance. At this point, we note that the lift force on intruders in \emph{static} beds, which exhibit a finite yield stress in the presence of gravity, has been studied extensively  \citep{wieghardt1974forces,ding2011drag,potiguar2013lift,guillard2014lift} and is well described by resistive force theory~\citep{ding2011drag,zhang2014effectiveness,agarwal2019modeling,agarwal2023mechanistic}. However, this approach is focused on static particle beds rather than the flowing granular media that we consider here.

In this paper, the lift force on a single spherical intruder particle in a dense, gravity-free linear shear flow is investigated using DEM simulations for a wide range of streamwise slip velocities. The net force on the intruder in the depthwise direction due to contacts with bed particles is defined as the lift force. Flow conditions (shear rate and overburden pressure) as well as intruder size are varied to consider a wide range of conditions. Particle scale analysis, including the local packing fraction, pressure field, and contact force distribution on the intruder, provides insight into the granular lift force. In addition, computational fluid dynamics (CFD) simulations of the analogous fluid flow over a spherical intruder are used to better understand the origin and differences of the lift force in granular flows relative to the lift force (Saffman effect) in fluid shear flows.
 
\section{Methods}\label{sec:methods}

\begin{figure}
  \begin{center}
  \includegraphics{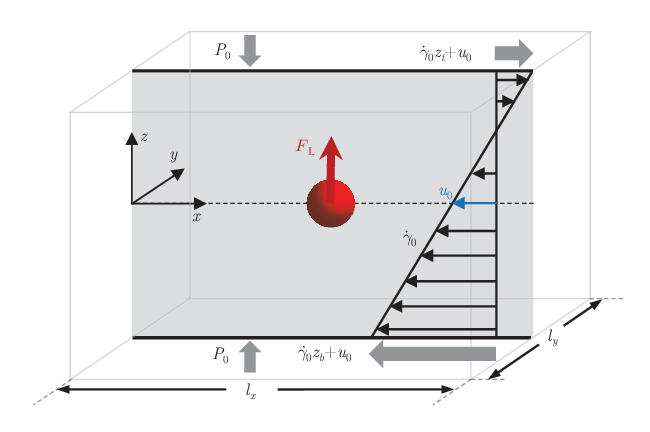} 
  \end{center}
  \caption{Schematic of the simulated zero-gravity linear shear flow (not to scale) with streamwise ($x$) velocity $u_0$ at the horizontal centerline used to characterize the lift force, $F_\mathrm{L},$ on a spherical intruder fixed at the center of the computational domain.}
  \label{setup}
\end{figure}

We consider a fixed intruder in gravity-free uniform shear flow, shown in figure~\ref{setup}.  We simulate the system using LIGGGHTS, an open-source DEM code, and apply periodic boundaries in the streamwise ($x$) and spanwise ($y$) directions. Two rough horizontal walls consisting of randomly distributed particles confine the flow at the top ($z_\mathrm{t}$) and bottom ($z_\mathrm{b}$), yielding an $l_x \times l_y \times l_z=z_\mathrm{t}-z_\mathrm{b}$ computational domain with a single intruder sphere fixed in the middle of the domain at $(x_\mathrm{i},y_\mathrm{i},z_\mathrm{i})=(l_x/2,0,0)$. The walls translate in the $x$ direction with a bottom wall velocity of $\dot\gamma_0 z_\mathrm{b}+u_0$ and a top wall velocity of $\dot\gamma_0 z_\mathrm{t}+u_0$. The prescribed wall motion imposes a uniform shear rate of $\dot\gamma_0$ and a mean streamwise velocity of $u_0$ (negative in figure~\ref{setup}) at $z=z_\mathrm{i}=0$ in the absence of an intruder. Following the usual convention~\citep{stone2000philip}, the slip velocity of the intruder relative to the flow is defined as $u_\mathrm{s}=u_\mathrm{i} - (\dot{\gamma}_0 z_\mathrm{i}+u_0)$, where $u_\mathrm{i}$ is the intruder velocity. Here we set $u_\mathrm{i}=0$ without loss of generality so that $u_\mathrm{s}=-u_0$.

During the initial 2\,s of each simulation, equal and opposite depthwise forces are applied to the shearing top and bottom walls to generate an overburden pressure, $P_0$. After this initial phase, both walls are fixed in the $z$-direction to set up a volume-constrained flow with $z_{\mathrm{t}}\approx -z_{\mathrm{b}}$. The fixed-volume domain simplifies comparison of our granular simulations to the corresponding fluid flow simulations and avoids fluctuations in the height of the granular bed due to shear-driven variations in the packing fraction that occur in the pressure-controlled approach. For the fixed bed-volume phase of the simulation, the time-averaged overburden pressure varies by less than 1\% from the value of $P_0$ applied during the initial constant overburden pressure phase of simulation. After stabilization of the fixed-volume flow for 2\,s, the average lift force, $F_\mathrm{L}$, is measured as the net vertical force on the intruder due to contacts with bed particles over 4\,s (corresponding to ten to forty shear rate time scales, $1/\dot\gamma_0$, for the range of $\dot\gamma_0$ tested). A comparison between this direct measurement method of the net force to the virtual spring method~\citep{guillard2016scaling} is provided in Appendix~\ref{app_comparison_F_L_methods} and demonstrates that the two methods yield nearly identical results, as similarly noted by \citet{van2018segregation}. The direct measurement method is preferred here because the intruder remains at a fixed vertical position such that $u_\mathrm{s}$ is constant.

The lift force is sensitive to the distance of the intruder from the upper and lower walls as well as to the lengths of the two periodic dimensions due to interactions with mirrored images of the intruder particle. Based on extensive testing to determine when the lift force is independent of the domain size, we set $l_x =10 d_\mathrm{i}$ (streamwise), $l_y =10 d_\mathrm{i}$ (spanwise) and $l_z \approx54 d_\mathrm{i}$ (depthwise), where $d_\mathrm{i}$ is the intruder diameter (unless otherwise noted). The number of particles in each simulation ranges from approximately 5000 to 350000, depending on the intruder to bed particle size ratio, $R=d_\mathrm{i}/d$, where $d=5$\,mm is the mean bed particle diameter with $\pm10\%$ uniform polydispersity and $d_\mathrm{i}$ is changed to vary $R \in \{1, 2, 3, 4\}$.  The intruder density equals the bed particle density, $\rho=2500$\,kg/m$^3$. 

To eliminate forces induced by shear rate gradients in the base flow~\citep{jing2021unified}, a uniform linear velocity profile is imposed by applying a small streamwise stabilizing force, $K_\mathrm{s}(\dot\gamma_0 z_\mathrm{p}-u_\mathrm{p})$, at every time step to all bed particles with centers more than $3 d_\mathrm{i}$ away from the center of the intruder, where $z_p$ and $u_p$ are the instantaneous vertical position and streamwise velocity of a bed particle, respectively, and $K_\mathrm{s}$ is a constant \citep{lerner2012unified,clark2018critical,fry2018effect,duan2020segregation,jing2022drag}. Based on a scaling analysis~\citep{jing2022drag}, the dimensionless version of this parameter is set to $\tilde{K}_\mathrm{s}=K_\mathrm{s}\dot\gamma_0 I^{-0.25}P_0^{-1} d^{-1}=0.1$ for all simulations, where $I=\dot\gamma_0d/\sqrt{P_0/\rho}$ is the inertial number. This small additional forcing assures that a linear velocity profile is maintained away from the intruder while not affecting the granular flow rheology. The spherical volume of radius $3 d_\mathrm{i}$ surrounding the intruder where the velocity control scheme is not applied ensures that the controller does not affect local interactions of the bed particle and the intruder, which, otherwise, could alter the measured lift force. 

We focus on dense granular flows ($0.01 \leq I \leq 0.5$) \citep{jop2015rheological}, with applied overburden pressures $P_0 \in \{250, 500, 1000, 2000\}$\,Pa, and imposed shear rates $\dot\gamma_0 \in \{2.5, 5, 10\}$\,s$^{-1}$. The shear rate is spatially uniform away from the intruder and there is no gravity in order to avoid forces induced by shear rate gradients and buoyancy, respectively~\citep{jing2021unified}. The DEM simulations use the Hertzian contact model with Young's Modulus $E=5\times10^7$\,Pa, Poisson's ratio $\nu=0.4$, particle-particle friction coefficient $\mu_p=0.5$, and restitution coefficient $e=0.8$. A sensitivity analysis that changes one parameter at a time and keeps the other parameters at their default values indicates that the results are insensitive to the precise values of these parameters in the tested ranges of $E \in \{10^{7}, 10^{9}\}$\,Pa, $\nu \in \{0.3, 0.5\}$, $\mu_{p} \in \{0.3, 0.7\}$, and $e \in \{0.6, 0.99\}$. The simulation time step is $\Delta t=10^{-5}$\,s for computational stability.

\section{Results}\label{sec:results_discussions}
  
\subsection{Lift force}
\subsubsection{General response}\label{Typical_case}

\begin{figure}
  \centerline{\includegraphics{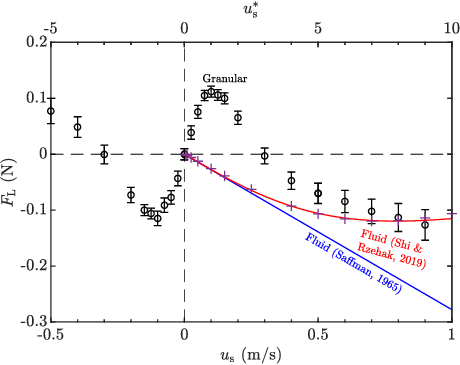}}
  \caption{Variation of intruder lift force, $F_\mathrm{L},$ with slip velocity, $u_\mathrm{s},$ (bottom axis) and non-dimensionalized slip velocity, $u^*_\mathrm{s}=u_\mathrm{s}/(\dot\gamma_0(d_\mathrm{i}+d)),$ (top axis)  in a granular shear flow for size ratio $R=3$, overburden pressure $P_0= 1000$\,Pa, and shear rate $\dot\gamma_0=5$\,s$^{-1}$ from DEM simulations ($\circ$). Error bars indicate standard error of $F_\mathrm{L}$ considering temporal correlations \citep{zhang2006calculation}. Solid blue and red curves indicate predicted lift forces in comparable fluid shear flows at low $Re_\mathrm{i}$, see text. Results from fluid CFD simulations ($+$) match predictions of \citet{shi2019lift}, as discussed in Sec.~\ref{Fluid_simulation}.}
  \label{F_LvsVsdimensional}
\end{figure}

Figure~\ref{F_LvsVsdimensional} shows the lift force, $F_\mathrm{L},$ on a non-rotating spherical intruder in a uniform granular shear flow as a function of the slip velocity, $u_\mathrm{s}$, for $R=3$ ($d_\mathrm{i} = 1.5$\,cm), $P_0= 1000$\,Pa, and ${\dot{\gamma_0}}=5$\,s$^{-1}$. $F_\mathrm{L}$ is antisymmetric about $u_\mathrm{s}=0$, i.e.\ $F_\mathrm{L}(u_\mathrm{s}) = -F_\mathrm{L}(-u_\mathrm{s})$, as expected.  Consequently, for the remainder of this paper, we only consider $u_\mathrm{s} \geq 0$. Of primary interest is the non-monotonic dependence of $F_\mathrm{L}$ on $u_\mathrm{s}$: for $0 < u_\mathrm{s} \lesssim 0.1$\,m/s, $F_\mathrm{L}$ increases until it reaches a maximum, but then decreases monotonically for $u_\mathrm{s} > 0.1$\,m/s, changing sign at $u_\mathrm{s} \approx 0.3$ \,m/s. Also shown in the figure are theoretical predictions of the lift force on a non-rotating spherical intruder in a uniform fluid shear flow by Saffman~\citep{saffman1965lift} as well as from a subsequent and improved calculation~\citep{shi2019lift}, where the fluid kinematic viscosity is 0.0536~m$^2$/s in order to match the granular viscosity determined by the $\mu(I)$ rheology as described in Sec.~\ref{Fluid_simulation}.  Surprisingly, the direction of $F_\mathrm{L}$ for an intruder in a granular shear flow is opposite that for the Saffman lift force in a fluid at lower slip velocities ($0 < u_\mathrm{s} \lesssim 0.3$\,m/s) as first noted by \citet{yennemadi2023drag}. For $u_\mathrm{s} > 0.3$\,m/s, $F_\mathrm{L}$ acts in the same direction as the fluid lift force. The granular lift force exceeds the fluid lift force predicted by \citet{shi2019lift} above $u_\mathrm{s}\approx0.7$\,m/s, but we do not verify whether it too eventually again decreases in magnitude at larger $u_\mathrm{s}$ as in the model of \citet{shi2019lift} due to the size of the computational domain necessary for simulations at larger values of $u_\mathrm{s}$. 

We non-dimensionalize the slip velocity as $u^*_\mathrm{s}=u_\mathrm{s}/(\dot\gamma_0(d_\mathrm{i}+d))$, where the velocity scale $\dot\gamma_0(d_\mathrm{i}+d)$ captures the shear-induced velocity difference across the intruder based on the undisturbed velocity of contacting bed particles immediately above and below the intruder. The maximum lift force occurs at $u^*_\mathrm{s} \approx 1$ (as indicated on the horizontal axis at the top of figure~\ref{F_LvsVsdimensional}), which corresponds to relative bed particle velocities of $-\dot\gamma_0(d_\mathrm{i}+d)/2$  just above the intruder and $-{3\dot{\gamma}}_0(d_\mathrm{i}+d)/2$ just below the intruder. $F_\mathrm{L}$ changes sign at $u^*_\mathrm{s} \approx 3$ with corresponding relative bed particle velocities of $-5\dot\gamma_0(d_\mathrm{i}+d)/2$ above the intruder and $-{7\dot{\gamma}}_0(d_\mathrm{i}+d)/2$ below the intruder. The remainder of this paper explores the scaling of $F_\mathrm{L}$ and the conspicuous differences between $F_\mathrm{L}$ in granular and fluid shear flows. 

\begin{figure}
  \centerline{\includegraphics{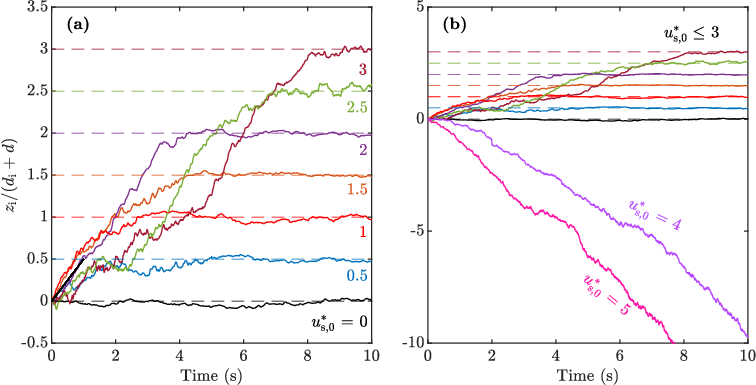}}
  \caption{Depthwise trajectories of free intruder for various initial non-dimensionalized slip velocities $u^*_{\mathrm{s},0}$: (a) $0 \leq u_{\mathrm{s},0}^* \le 3$, and (b) $0 \leq u_{\mathrm{s},0}^* \le 5$ with $R=3$, $P_0=1000$\,Pa and $\dot\gamma_0=5$\,s$^{-1}$.  When $F_\mathrm{L}<0$ (for $u_{\mathrm{s},0}^* = 4$ or $5$ here, see figure~\ref{F_LvsVsdimensional}), the intruder moves downward until it reaches the vicinity of the lower wall.
  }
   \label{z-track}
\end{figure}

To verify the lift force results shown in figure~\ref{F_LvsVsdimensional}, we examine the depthwise ($z$-axis) trajectory of an intruder that is unconstrained in the $z$-direction for different fixed values of the bed velocity at the vertical centerline, $u_0,$ corresponding to different initial values of the non-dimensionalized slip velocity, $u^*_{\mathrm{s},0}$. The intruder starts at $z=0$ with an initial slip velocity of $u_\mathrm{s}=-u_0$, but as time progresses, the intruder should rise or sink due to a non-zero lift force. Results for an intruder with $R=3$ are shown in figure~\ref{z-track}(a). For $u_0=0$, where the lift force is zero, the intruder neither rises nor sinks, corresponding to the case where the bed particle velocities above and below the intruder are equal and opposite.  For $0 < u^*_{\mathrm{s},0} < 3$, where the initial lift force is positive, the intruder moves upward, as shown in figure~\ref{z-track}(a), but then stops moving upward when it reaches a position where it has the same streamwise velocity as the local bed particles, i.e.\ $u_\mathrm{s}=0$ at $z_\mathrm{i}/(d_\mathrm{i}+d) = u^*_{\mathrm{s},0}$.  We do not quantify the upward velocity of the intruder in this regime due to the relatively large collision-induced fluctuations, but it is evident that the upward velocities are similar for different $u_\mathrm{s}$ and tend to slow as the intruder reaches its steady-state position.  
However, for $u^*_{\mathrm{s},0} = 3$, where the initial lift force is approximately zero, the intruder does not rise for the first half-second. Nevertheless, once the intruder has risen slightly, its instantaneous slip velocity decreases, resulting in a positive and increasing lift force that eventually carries the intruder to its equilibrium $z$-position at $z_\mathrm{i}/(d_\mathrm{i} + d)=3$.

A very different trajectory occurs when $u^*_{\mathrm{s},0} > 3$, where the initial lift force is negative, as shown in figure~\ref{z-track}(b) which has different vertical scale limits than in (a) and includes the cases for $u^*_{\mathrm{s},0} \leq 3$ for comparison. When $F_\mathrm{L} < 0 $, the intruder moves downward faster than the intruder rises when $F_\mathrm{L} >0$. As the intruder moves downward, the slip velocity increases and the lift force becomes increasingly negative as $u^*_\mathrm{s}$ increases (recall figure~\ref{F_LvsVsdimensional}). Although not shown in the figure, the intruder in these cases continues moving downward with increasing speed until it nearly reaches the bottom wall. The main point to be taken from figure~\ref{z-track} is that the free intruder trajectories are consistent with the lift force measurements for a vertically tethered intruder, shown in figure~\ref{F_LvsVsdimensional}, thereby confirming the surprising result of positive and non-monotonic lift at small $u^*_{\mathrm{s},0}$, which is opposite to that which occurs in the fluid shear case at small $u^*_{\mathrm{s},0}$.

\subsubsection{Lift force scaling}\label{dependence_parameters}
\begin{figure}
  \center{\includegraphics{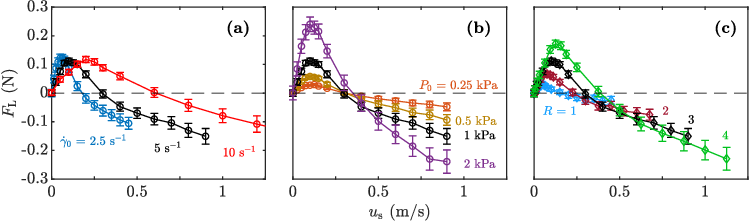}}
  \caption{$F_\mathrm{L}$ vs.\ $u_\mathrm{s}$ for a non-rotating intruder with varying (a) $\dot{\gamma}_0$ with $R= 3$, $P_0= 1000$ Pa; 
  (b) $P_0$ with $R= 3$, $\dot{\gamma}_0 = 5$\,s$^{-1}$; and (c) $R$ with $P_0= 1000$ Pa, $\dot{\gamma}_0 = 5$\,s$^{-1}.$ Note that two data points at large $u_\mathrm{s}>1.25$\,m/s for $\dot{\gamma}_0 = 10$\,s in (a) are not shown to keep the horizontal axis consistent in all three plots. 
  }
  \label{F_l_par_varied}
\end{figure}

To understand the dependence of the lift force on flow conditions, we vary the shear rate $\dot\gamma_0$, the overburden pressure $P_0$, and the particle size ratio $R$. Simulations are limited to $u_\mathrm{s}^* \leq 9$, above which computations are increasingly more difficult due to the large computational domain that is necessary. Figure~\ref{F_l_par_varied}(a) demonstrates how the relation for $F_\mathrm{L}$ versus $u_\mathrm{s}$ depends on $\dot\gamma_0$. All three curves exhibit the same non-monotonic trend with increasing $u_\mathrm{s}$, and their peak values of $F_\mathrm{L}$ are similar. However, with increasing $\dot\gamma_0$, the maximum and zero-crossing points of $F_\mathrm{L}$ both shift to larger values of $u_\mathrm{s}$, indicating that $F_\mathrm{L}$ is linked to the intruder slip velocity relative to the bed shear rate rather than to the absolute intruder slip velocity.

Varying the overburden pressure $P_0$ has a different impact on $F_\mathrm{L}(u_\mathrm{s})$, as shown in figure~\ref{F_l_par_varied}(b). Although the curves follow the same non-monotonic trend, reaching their peaks and decreasing to zero at similar values of $u_\mathrm{s}$, the magnitude of $F_\mathrm{L}$ increases with increasing $P_0$. This increase in $F_\mathrm{L}$ with $P_0$ is not a consequence of closer bed particle packing at higher pressures, since packing density $\phi \approx 0.58$ regardless of overburden pressure.  Instead, it appears to be due to the proportionality of overburden pressure and stresses in the granular bed. 

Figure~\ref{F_l_par_varied}(c) shows $F_\mathrm{L}(u_\mathrm{s})$ for different values of $R$. The same non-monotonic trend occurs with different $R$, but the $F_\mathrm{L}$ maximum and the zero-crossing point both shift to larger $u_\mathrm{s}$ values with increasing $R$, suggesting that the slip velocity scales with the intruder size. In addition, the magnitude of $F_\mathrm{L}$ increases with increasing $R$, which is reasonable given that both the velocity difference across the intruder and the intruder surface area increase as the intruder diameter is increased.

\begin{figure}
  \centerline{\includegraphics{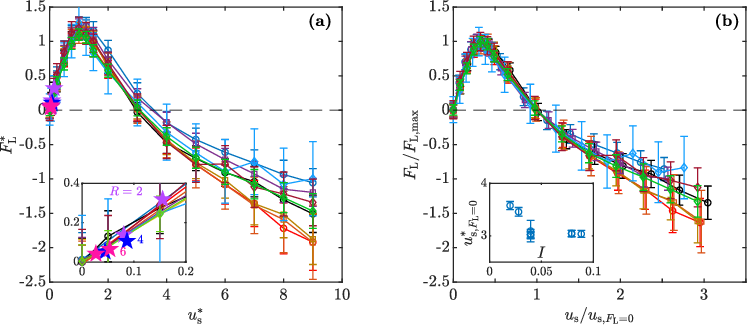}}
  \caption{Scaled lift force vs.\ scaled slip velocity for all data in figure~\ref{F_l_par_varied}.  (a) $F^*_\mathrm{L}$ vs.\ $u^*_\mathrm{s}$, where the symbols are the same as those in figure~\ref{F_l_par_varied}. Pentagram symbols indicate data from \citet{yennemadi2023drag} for an intruder in linear shear flow with $R\in \{2, 4, 6\}$.  (b) $F_\mathrm{L}/F_\mathrm{L,max}$ vs.\ $u_\mathrm{s}/u_{\mathrm{s},F_\mathrm{L}=0}$, where $F_\mathrm{L,max}$ and $u_{\mathrm{s},F_\mathrm{L}=0}$ indicate the peak force and slip velocity where $F_\mathrm{L}=0$, respectively. Inset:  $u^{*}_{\mathrm{s},F_\mathrm{L}=0}$ vs.\ $I$.}
  \label{F_l_scaled_par_varied}
\end{figure}

The qualitatively similar dependence of $F_\mathrm{L}$ on $u_\mathrm{s}$ over a range of parameter values as illustrated in figure~\ref{F_l_par_varied} suggests that the results may collapse when scaled appropriately. Accordingly, we non-dimensionalize the lift force as $F^*_\mathrm{L}=F_\mathrm{L}/\left(P_0(d_\mathrm{i}+d)^2/4\right)$, since $P_0(d_\mathrm{i}+d)^{2}/4$ is proportional to the intrinsic contact force scale in this dense linear shear flow, and plot it versus the scaled slip velocity $u^*_\mathrm{s}=u_\mathrm{s}/(\dot\gamma_0(d_\mathrm{i}+d))$. 
This simple scaling is surprisingly effective as shown in figure~\ref{F_l_scaled_par_varied}(a), which includes all the data sets from figure~\ref{F_l_par_varied}. All of the $F^*_\mathrm{L}(u^*_\mathrm{s})$ curves collapse for $u^*_\mathrm{s}\leq 1$.
Furthermore, the maximum in $F^*_\mathrm{L}$ occurs at $u^*_\mathrm{s} \approx 1$, suggesting that the slip velocity scaling is not only dimensionally correct, but that the velocity scale $\dot{\gamma}_0 (d_\mathrm{i}+d)$ reflects the magnitude of the slip velocity for maximal lift. Likewise, the maximum value for the lift force of $F^*_\mathrm{L}\approx 1$ indicates that the intrinsic contact force scale of $P_0(d_\mathrm{i}+d)^{2}/4$ is appropriate. The collapse is less satisfying for $u^*_\mathrm{s} > 1$, although the spread is surprisingly small given the range of values for $\dot{\gamma}_0$, $P_0$, and $R$ included in the plot.

While the scaled-data show good collapse in figure~\ref{F_l_scaled_par_varied}(a), the maximum lift force, $F_\mathrm{L,max}$, and the slip velocity where $F_\mathrm{L}=0$ at finite slip velocity, $u_{\mathrm{s},F_\mathrm{L}=0}$, vary slightly between parameter sets. To more directly test if $F_\mathrm{L}(u_\mathrm{s})$ is self-similar and to further validate the proposed scaling, we alternatively scale the slip velocity by $u_{\mathrm{s}, F_\mathrm{L}=0}$ and the lift force by $F_\mathrm{L,max}$, where, for each data set, $u_{\mathrm{s}, F_\mathrm{L}=0}$ is obtained from a linear fit of $F_\mathrm{L}(u_\mathrm{s})$ about the zero crossing and $F_\mathrm{L,max}$ is obtained from a parabolic fit about the maximum. The result, shown in figure~\ref{F_l_scaled_par_varied}(b), demonstrates excellent collapse, thereby confirming that $F_\mathrm{L}(u_\mathrm{s})$ is indeed self-similar for the parameter range this work explores.  The greater variation of the collapsed data in figure~\ref{F_l_scaled_par_varied}(a) compared to figure~\ref{F_l_scaled_par_varied}(b) is due to the increase in $u^{*}_{\mathrm{s},F_\mathrm{L}=0}$ for conditions with the lowest inertial numbers, namely, $\{2.5\,\mathrm{s}^{-1}, 1000\,\mathrm{Pa}, 3\}$ and $\{5\,\mathrm{s}^{-1}, 2000\,\mathrm{Pa}, 3\}$, for which $I=0.020$ and $I=0.028$, respectively. The inset of figure~\ref{F_l_scaled_par_varied}(b) shows that $u^{*}_{\mathrm{s},F_\mathrm{L}=0} = u_{\mathrm{s},F_\mathrm{L}=0}/\left(\dot\gamma_0(d_\mathrm{i}+d)\right)$ increases with decreasing inertia number for $I<0.03$, a range approaching the lower end of the dense granular flow regime, $0.01 \leq I \leq 0.5$~\citep{jop2015rheological} where the flow approaches that quasistatic flow regime.


In a previous study, 
\citet{yennemadi2023drag} directly measure the lift force for $R\in\{2, 4, 6\}$ for linear shear flow with $u^*_\mathrm{s} < 0.2$, a much smaller range of scaled slip velocities than explored here.  Their data (pentagram symbols in figure~\ref{F_l_scaled_par_varied}(a) and its inset) are consistent with our results and scaling. In another study, \citet{van2018segregation} indirectly estimate the lift force in gravity-driven chute flow with even smaller slip velocities, $\left|u^*_\mathrm{s}\right|\leq 0.01$. They find a much larger lift force with the opposite sign compared to our results and those of \citet{yennemadi2023drag}. We believe this can be attributed to their assumptions made to estimate the buoyancy force, as explained in Appendix~\ref{Saffman}.

\subsection{Drag}
\label{sec:drag}
The intruder lift force is determined by the depthwise component of the integral of the normal and tangential stresses acting on the intruder surface.  Similarly, the drag force equals the component of the same surface integral in the direction opposite to the intruder slip velocity.  Given the common dependence of both forces on the surface stresses,  we now examine the drag force on the spherical intruder as it slips relative to the flow in the streamwise direction as well as the lift-to-drag ratio, which highlights the differing dependence of the two forces on the slip velocity.  The drag force is measured directly as the net streamwise force on the intruder due to contacts with bed particles and is typically averaged over 4\,s (corresponding to $10/\dot{\gamma}_0$ to $40/\dot{\gamma}_0$). The results are recast into the form of a drag coefficient, $C_\mathrm{D}=8|F_\mathrm{D}|/({\rho}{\pi}d^{2}_{\mathrm{i}}u^{2}_{\mathrm{s}})$, and its dependence on the intruder Reynolds number, $Re_\mathrm{i}={\rho}d_\mathrm{i}u_\mathrm{s}/{\eta}$. The viscosity, $\eta$, is based on the granular $\mu(I)$ rheology \citep{jop2006constitutive}, where $\mu(I)=\frac{\tau}{P_0}$ is the stress ratio (also called the macroscopic friction coefficient~\citep{dumont2023microscopic}) in the flow. The viscosity is estimated as $\eta = \mu(I) \frac{P_0}{\dot{\gamma}_0}$, where $\mu(I) = \mu_\mathrm{s} +\frac{\mu_2-\mu_\mathrm{s}}{I_\mathrm{c}/I+1} $ with $\mu_\mathrm{s} = 0.36$, $\mu_2 = 0.91$, and $I_c = 0.73$ for simple shear flow using identical bed particles~\citep{jing2022drag}.

\begin{figure}
  \centerline{\includegraphics{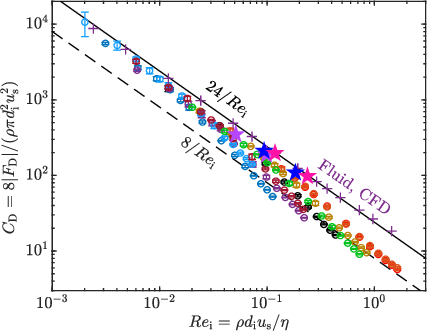}}
  \caption{Drag coefficient, $C_\mathrm{D},$ vs.\ intruder Reynolds number, $Re_{\mathrm{i}},$ for the parameter value combinations in figure~\ref{F_l_par_varied}. The CFD drag results ($+$) are discussed in Section~\ref{Fluid_simulation}. Symbols are the same as in figures~\ref{F_l_par_varied} and~\ref{F_l_scaled_par_varied}.}
  \label{Drag_coeff}
\end{figure}

Figure~\ref{Drag_coeff} shows a general Stokes-like linear relationship between $C_\mathrm{D}$ and $Re_\mathrm{i}$ with $8/Re_\mathrm{i} \lesssim C_\mathrm{D} \lesssim 24/Re_\mathrm{i}$ over all simulations, indicating that the drag force is proportional to the slip velocity. This result is important in several ways. First, the result is consistent with previous extensive measurements showing that the drag force on an intruder in a granular shear flow is Stokes-like~\citep{jing2022drag, tripathi2011numerical}. Second, and perhaps more interesting, is that previous drag results apply to an intruder moving normal to the shear flow (across flow streamlines), while our results are for an intruder moving in the streamwise direction of the shear flow. That the drag is consistent between these two scenarios suggests universality of the Stokes drag relation for spherical intruders in sheared granular flows. Third, the drag force measured here not only follows Stokes' law, but it is also nearly within the range found previously, $8/Re_\mathrm{i} \le C_\mathrm{D} \le 24/Re_\mathrm{i}$ for $10^{-6} \le Re_\mathrm{i} \le 10^1$~\citep{jing2022drag}, indicating that the bounds associated with granular Stokes drag are likely broadly applicable to other types of flow and intruder slip directions relative to those flows.  Note, however, that the slope of $C_\mathrm{D}$ versus $Re_\mathrm{i}$ is less than $-1$ at higher $Re_\mathrm{i}$ values in each data set and that the scaling of $C_\mathrm{D}$ changes from $C_\mathrm{D}\approx 20/Re_\mathrm{i}$ at lower $Re_\mathrm{i}$ and appears to approach the limiting form drag relation of $C_\mathrm{D}\approx 8/Re_\mathrm{i}$ at larger $Re_\mathrm{i}$.  This change suggests that at large slip velocities, pressure dominates friction based on the analogy to drag on an intruder in a viscous fluid flow having a value of $8/Re_\mathrm{i}$ for the form (pressure) contribution to the overall drag.

\begin{figure}
  \centerline{\includegraphics{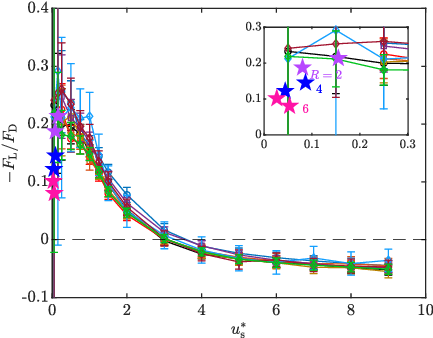}}
  \caption{Lift-to-drag ratio, $-F_\mathrm{L}/F_\mathrm{D},$ vs.\ $u^*_\mathrm{s}$ for varying parameters (symbols and colors) as indicated in figures~\ref{F_l_par_varied} and~\ref{F_l_scaled_par_varied}. Pentagram symbols are from ~\citet{yennemadi2023drag} for an intruder in linear shear flow and $R\in\{2,4,6\}$~\citep{yennemadi2023drag}. Inset: low $u^*_\mathrm{s}$ data.  
 }
  \label{Lift_drag_ratio}
\end{figure}

As mentioned above, the lift and drag forces are the depthwise and streamwise components of the integrated surface stresses acting on the intruder. If these two forces were simply proportional to the slip velocity, the lift-to-drag ratio, $-F_\mathrm{L}/F_\mathrm{D}$ (note that $F_\mathrm{D}$ is negative as shown in figure~\ref{setup}), would be independent of $u^*_\mathrm{s}.$  However, this is not the case, as figure~\ref{Lift_drag_ratio} shows.  The data for different simulation parameters collapse reasonably well, but $-F_\mathrm{L}/F_\mathrm{D}$ is obviously not constant except, perhaps, for $u^*_\mathrm{s}<0.3$ where $0.2 < -F_\mathrm{L}/F_\mathrm{D} < 0.3$ for the different data sets (see inset). 
These ratios are somewhat higher than previous results with low $u^*_\mathrm{s}$ \citep{yennemadi2023drag}, particularly for $R=4$ and $6$, because their drag forces are higher due to higher values of $I=0.23$ and $R$ up to 6, in agreement with \citet{jing2022drag}.
For increasing $u^*_\mathrm{s}$,  $-F_\mathrm{L}/F_\mathrm{D}$ decreases with decreasing slope magnitude, becomes negative when $F_\mathrm{L}$ reverses to match the Saffman lift direction in a fluid, and appears to approach an asymptotic value of $-F_\mathrm{L}/F_\mathrm{D}\approx -0.03$ for large $u^*_\mathrm{s}$. In all cases, the lift force magnitude is about one order of magnitude smaller than the drag force magnitude over the range of $u^*_\mathrm{s}$ that we consider.

\subsection{Fluid lift comparison} \label{Fluid_simulation}
To gain further insight into the lift force on a spherical intruder in granular shear flow, we simulate the analogous problem in a fluid and compare and contrast how shear and pressure contribute to lift in both cases. Of course, many factors contribute to the lift force in a fluid (see Appendix~\ref{Fluid_lift}). Nevertheless, the point of this fluid comparison is to examine how the constitutive differences between granular flow and incompressible Newtonian flow contribute to the different lift forces evident in figure~\ref{F_LvsVsdimensional}.
The incompressible Newtonian fluid simulations are implemented in ANSYS Fluent, a commercial CFD simulation package, for the same flow geometry as that shown in figure~\ref{setup}. The intruder and fluid parameters are matched to the DEM simulations in Sec.~\ref{Typical_case} for $R=3$: $d_\mathrm{i}=1.5$\,cm, fluid density $\rho_f=1450$\,kg/m$^3$, which equals the bed particle bulk density in granular flow simulations for $\phi=0.58$, and kinematic viscosity equal to $0.0536$\,m$^2$/s, which is based on the $\mu(I)$ rheology~\citep{jop2006constitutive} for $I=0.04$ as described in Sec.~\ref{sec:drag}. The intruder is modelled as a stationary non-rotating sphere with a no-slip boundary condition at its surface. The fluid velocities on all outer boundaries (see figure~\ref{setup}), i.e., upper and lower moving walls and the periodic inlet and outlet planes are prescribed~\citep{salem1998shear, mikulencak2004stationary} to produce a linear shear velocity profile in the $z$-direction with $\dot{\gamma}_0=\SI{5}{\per \second}$ and a streamwise fluid velocity of $-u_\mathrm{s}$ at $z=z_\mathrm{i}=0$ sufficiently far from the intruder. 

In incompressible Newtonian flow, shear stress is independent of absolute pressure so that an arbitrary background hydrostatic pressure can be superimposed on the pressure field. Here, the average background pressure on the outer boundaries is set to match the granular flow overburden pressure, $P_0= \SI{1000}{\pascal}$, for ease of comparison, and gravity is set to zero. Note that since we do not model the liquid-gas phase transition (i.e., cavitation) and the background pressure value is arbitrary with respect to the absolute pressure, the local fluid pressure can be negative.  
In the granular flow, however, the background pressure is physically relevant as it sets the scale of the lift and drag forces on the intruder in dense flows as we demonstrate above.  Additionally, the pressure cannot be negative in granular flow, since zero pressure represents a region devoid of bed particles for the non-cohesive particle flows we study here. 

Regarding the spatial discretization used in the solver,  the gradients of solution variables at cell centers are least-squares cell-based~\citep{barth19913}. Second order and second order upwind integration schemes are used for pressure and momentum interpolations, respectively. At convergence, the momentum and continuity equations have a maximum scaled residual of $10^{-12}$.  As in granular shear flow, the lift force in a fluid shear flow is sensitive to the domain size~\citep{salem1998shear,shi2019lift}, necessitating a domain size convergence study to ensure that the system size is sufficiently large to not affect the results. The domain size for which the lift force is unaffected by the bounding walls for the range of parameters and slip velocities investigated is $200d_\mathrm{i}\times200d_\mathrm{i}\times200d_\mathrm{i}$, consistent with results from a recent numerical study~\citep{andersson2019forces}, and much larger than the computational domain for the granular case ($10d_\mathrm{i}\times10d_\mathrm{i}\times54d_\mathrm{i}$).  Lift and drag forces are calculated by integrating the normal and tangential stresses on the intruder in the depthwise ($z$) and streamwise ($x$) directions, respectively. These forces agree with analytical and empirical results for the lift and drag forces \citep{stokes1851effect,saffman1965lift,shi2019lift} over the entire range of slip velocities considered in the DEM simulations, as shown in figures~\ref{F_LvsVsdimensional} and~\ref{Drag_coeff}, thereby confirming the validity of the fluid simulation results.



\begin{figure}
\centerline{\includegraphics[width=\textwidth]{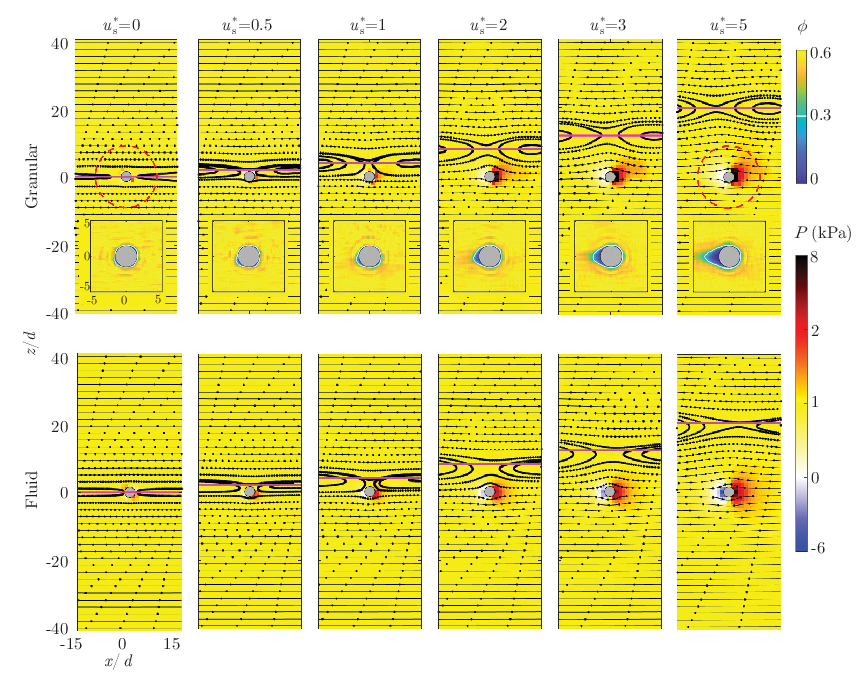}}
  \caption{Comparison of flow field pathlines for various scaled slip velocities, $u^*_\mathrm{s}$ in granular (top row) and fluid (bottom row) simulations averaged over time and a $d$-wide spanwise region centered on the intruder. Continuous color shading in the main panels indicates pressure, while color in the insets indicates the mean local packing density, $\phi$, around the intruder in granular flow. White in the pressure field indicates $P=0$, while the white packing fraction contour in the insets indicates $\phi =$ 0.3. The orange dashed circles represent the boundary of the controller-free region with radius 3\,$d_\mathrm{i}$ in the granular simulations (shown only in the first and last columns for reference). Flow conditions are $P_{0} =1000$\,Pa, $\dot\gamma_0=5$\,s$^{-1}$, and $R=3$.}
  \label{field}
\end{figure}

We first compare pathlines for granular and fluid shear flows based on the velocity field in an $xz$-plane centered on the intruder at various dimensionless slip velocities in figure~\ref{field}, where $d$ in the expression for $u^*_\mathrm{s}$ is set to $d=5$\,mm in the fluid case for consistency with the granular case. For the granular case, the average velocity field in a $d$-wide slice centered on the intruder is found using a binning procedure, and pathlines are determined by interpolating the average velocity in each bin. The horizontal magenta line indicates the location of zero velocity (matching the intruder velocity) at $z/d=(R+1)u^*_\mathrm{s}$, where the flow direction above the line is to the right and below the line is to the left. The velocity magnitude is indicated by the pathline dot spacing; dots are spaced at 0.05\,s time increments. The background in figure~\ref{field} indicates the local pressure, which equals the background pressure of $P_0=1000$\,Pa throughout most of the domain except near the intruder. Insets in the granular flow sub-figures (top row) indicate the local packing density, $\phi$, near the intruder. The orange dashed circles for two cases of granular flow indicate the boundary of the streamwise velocity controller-free region in all granular simulation cases. Note that the full computational domains for both granular and fluid systems extend vertically far beyond the figure limits to avoid wall effects.

It is immediately evident from figure~\ref{field} that the granular and fluid flow fields are quite similar. Fields and pathlines are symmetric above and below the intruder at $u^*_\mathrm{s}=0$ and asymmetric for $u^*_\mathrm{s}>0$ with much larger deviations in the pathlines above the intruder compared to below it and in the pressure in front compared to behind the intruder. One particularly noticeable feature is that instead of sweeping across the $xz$-plane horizontally, between one and three pathlines on the left and right side of the intruder cross the zero-velocity plane at $z/d=(R+1)u^*_\mathrm{s}$, reverse direction and never reach the pass the intruder center ($x=0$). For low slip velocities, the flow reverses just above the intruder and does not pass over the top of the intruder from right to left (evident for $u^*_\mathrm{s}\leq1$).  At higher $u^*_\mathrm{s}$ (for instance, $u^*_\mathrm{s}=2$) pathlines flow from right to left over the top of the intruder.  The situation where the flow does not pass over the intruder from right to left at low slip velocities has been noted for fluid flow~\citep{robertson1970low,poe1975closed} and is known as the blocking phenomenon~\citep{camassa2011lagrangian}. Increasing the slip velocity shifts the zero-velocity plane further above the intruder, thereby exposing the intruder to a higher right-to-left velocity in the shear field, thereby eliminating the blocking phenomenon.  

Intuitively, the transition from blocking to no blocking should occur at $u^*_\mathrm{s}=0.5$, when the zero-velocity plane is tangent to the top of the intruder. However, this ignores frictional effects. The frictional tangential boundary condition on a non-rotating intruder interferes with the intrinsic rotating component of the flow field, $\dot\gamma_0/2$, in linear shear flow. An additional set of fluid simulations with a free-slip boundary condition on the intruder confirms a transition from blocking to no blocking at $u^*_\mathrm{s}\approx0.5$ in the absence of friction. In contrast, in the fluid simulations with a no-slip boundary condition on the intruder and in DEM granular simulations with friction, the blocking phenomenon continues until $u^*_\mathrm{s}\approx 1$, as shown in figure~\ref{field}.  As a check, we note that \citet{camassa2011lagrangian} solved the flow field around a sphere in shear flow in the viscous limit and found that the transition from blocking to no blocking occurs at $u_\mathrm{s}\approx4\dot{\gamma}_0d_\mathrm{i}/3$. This corresponds to $u^*_\mathrm{s}\approx1$ for $R=3$, which is consistent with our results that the transition from blocking to no blocking occurs for $1\leq u_\mathrm{s}\leq2$, although we did not attempt to isolate the exact value of the transition.

Despite the pathline reversal just above the intruder, mass is conserved in both fluid and granular flows so that each granular or fluid element on the right side of the intruder that shifts from moving left to moving right is matched by a corresponding element on the left side of the intruder that shifts from moving right to moving left. In an infinitely long streamwise domain, the pathlines that change direction above the intruder would extend to negative infinity on the left and positive infinity on the right.  However, with the periodic streamwise boundary conditions used here, they instead form a horizontally elongated circulation roll, evident on the right side for $u_\mathrm{s}^*=5$ in the granular case. 

It is tempting to infer that lift forces result from the asymmetry of pathlines above and below the intruder, but this cannot be correct. Even though the pathlines are similar for the granular and fluid cases, the lift force has the opposite sign for the granular and fluid cases for $0<u_\mathrm{s}^*<3$. Nevertheless, there are subtle differences in the granular and fluid pathlines such as the slight shift of the point where the pathlines reverse direction above the intruder to the left in the granular case and to the right in the fluid case for $u_\mathrm{s}^*\geq2$.

The pressure fields around the intruder are also quite similar between the granular and fluid flow cases. Pressure is higher on the leading side (right side) of the intruder, most evident for large $u_\mathrm{s}^*$, due to the incoming right-to-left flow with respect to the intruder. The pressure appears nearly symmetric above and below the intruder, although the pressure is slightly higher above the intruder in the granular case for $u_\mathrm{s}^*\geq 2$ and slightly higher below the intruder for $u_\mathrm{s}^*=0.5,\,1$ in the fluid case. More importantly, the pressure field behind the intruder in the fluid case is negative for $u_\mathrm{s}^*\geq 2$, whereas the corresponding pressure in the granular case is zero since it cannot go lower. 

Lastly, it is helpful to examine the interactions of the bed particles with the intruder in the granular case. The packing density, $\phi$, shown in the insets in the top row of figure~\ref{field}, is near zero in a region behind the intruder for $u_\mathrm{s}^*\geq 2$. Thus, most of the particle contacts on the intruder occur on the leading side, while a void forms on the trailing side.
This corresponds to a constitutive difference between the granular shear flows, which sustains no tension on the trailing side of the intruder, and the fluid shear flows where the fluid remains in contact with the intruder on its trailing side and generates a negative pressure.

\begin{figure}  
\centerline{\includegraphics[width=\textwidth]{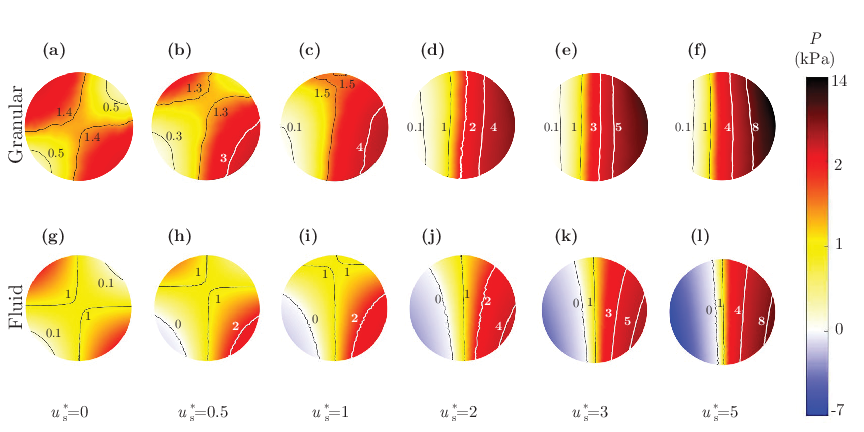}}
  \caption{Intruder pressure distribution (color map), $P$, in (a-f) granular DEM simulation and (g-l) fluid CFD simulation with $d_\mathrm{i}=1.5$\,cm.  See text for fluid simulation parameters. }
  \label{stress_dist}
\end{figure}

A more direct approach to understanding lift force differences between granular and fluid shear flows than considering the entire pressure field is to examine the pressure on the intruder surface that develops due to flow interactions. Figure~\ref{stress_dist} compares the pressure distributions on the intruder in granular DEM simulations (a-f) with fluid CFD simulations (g-l) at multiple slip velocities. The pressure distributions at $u^*_\mathrm{s}=0$ are similar in granular flow (a) and fluid flow (g), although the maximum pressure on the intruder in the granular case is higher. As the slip velocity increases, the pressure increases on the leading side of the intruder in both cases. However, there is a large difference in the pressure distribution on the trailing side for the two cases as is clearly evident in figures~\ref{stress_dist}(b-f) and (h-l). The pressure on the trailing side in the fluid case becomes negative (blue regions), and the negative pressure zone grows larger with increasing $u_\mathrm{s}^*$. In the granular case, the minimum pressure at large $u_\mathrm{s}^*$ on the trailing side of the intruder is zero, corresponding to the void evident in the insets in the top row of figure~\ref{field}.

\begin{figure}
  \centerline{\includegraphics[width=\textwidth]{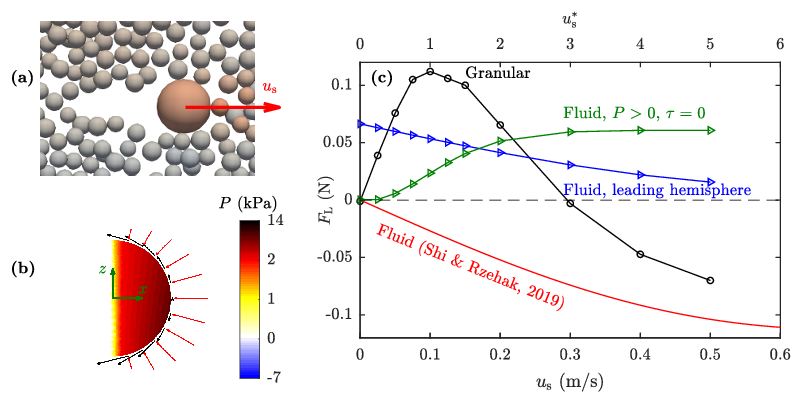}}
  \caption{Leading-hemisphere modification of the fluid results for $R=3$, $P_0= \SI{1000}{\pascal}$ and $\dot\gamma_0=\SI{5}{\per \second}$.
  (a) Snapshot of intruder and bed particles in DEM simulation in $y=0$ plane (centered on intruder) showing the void behind the intruder at dimensionless slip velocity $u^*_\mathrm{s}=5$.  (b) Pressure field, $P$, (color map) on the leading hemisphere of intruder in a fluid, where arrows are proportional to the normal (red) and tangential (black) stress components for $u^*_\mathrm{s}=5$ in the $y=0$ plane.
  (c) Lift force, $F_\mathrm{L},$ vs.\ slip velocity, $u_\mathrm{s},$ for granular flow simulation (black),  fluid theory (red), and CFD fluid simulation with leading-hemisphere modification (blue) and $(P>0,\tau=0)$-modification (green). 
  }
  \label{modification1}
\end{figure}

The qualitative similarity between the pressure distributions on the leading side of the intruder in granular and fluid shear flows suggests that it might be possible to artificially account for the constitutive differences between the fluid and granular media by simply considering only the contributions to the lift force on the leading side of the intruder.  This eliminates the differences between a tractionless void in the granular case, shown in a snapshot from a DEM simulation in figure~\ref{modification1}(a), and the negative pressure in the fluid case on the trailing side of the intruder.  In other words, including only the stresses on the leading (right) hemisphere of the intruder in the fluid case (\textquotesingle leading-hemisphere\textquotesingle~modification) might be a simple way to modify the fluid case to mimic the granular case. Figure~\ref{modification1}(b) shows the fluid pressure distribution on the leading hemisphere of the intruder in the fluid case as well as vectors corresponding to the normal (red) and tangential (black) components of the stress for $u_\mathrm{s}^*=5$. The associated lift force calculated using only the leading hemisphere in the fluid case is shown in figure~\ref{modification1}(c) versus $u_\mathrm{s}$ (bottom horizontal axis) and $u^*_\mathrm{s}$ (top horizontal axis). The sign of the lift force for the modified fluid case flips so that $F_\mathrm{L}$ is in the same direction and similar in magnitude to $F_\mathrm{L}$ in the granular case. This suggests that the void on the trailing side of the intruder in granular flow is at least partially responsible for the opposite sign of the granular lift force compared to the fluid case. However, this is clearly only part of the story. First, this approach results in a finite value for $F_\mathrm{L}$ at $u_\mathrm{s}=0$, which is physically incorrect.  Second,  $F_\mathrm{L}$ based on the leading-hemisphere-only modification does not become negative as $u_\mathrm{s}$ increases, even at large slip velocities (we simulated values as large as $u_\mathrm{s}^*=45$, not shown in figure~\ref{modification1}(c)).  To fully understand the differences between the two cases, a more sophisticated analysis of the $z$-components of the stresses on the intruder is necessary.

\subsection{Vertical stress}
\label{stress_distribution}
Based on the results in the previous section, it is clear that differences in the stress distributions on the surface of the intruder determine both the sign and magnitude of the lift force. Here we analyze the lift direction component of the combined normal stress and shear stress as well as of each stress component individually to understand their influences on $F_\mathrm{L}$.

\begin{figure}
\centerline{\includegraphics[width=\textwidth]{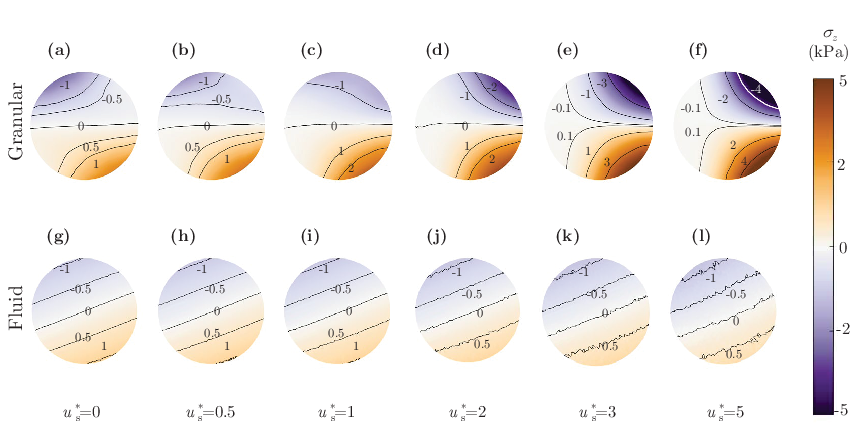}}
  \caption{Net vertical stress distribution on the intruder ($\sigma_{z}$) in (a-f) granular flow and (g-l) fluid flow with increasing (left to right) dimensionless slip velocity, $u^*_\mathrm{s}$. Positive stresses (orange) act upward and negative stresses (purple) act downward. Zero stress contours are omitted in some cases for clarity.}
  \label{sigmaz}
\end{figure}

First consider the lift-force component of the combined normal and shear stress fields acting on the intruder surface, $\sigma_z$, in both granular ($R=3$) and fluid flow, shown in figure~\ref{sigmaz}. The vertical stress distributions differ substantially between the granular and fluid cases over the range of slip velocities. For both flows, when $u^*_\mathrm{s}=0$ the downward stress (lavender) zone on the top-left of the intruder and the upward stress (orange) zone on the bottom-right of the intruder are approximately symmetric about the intruder center. This is reasonable because for $u^*_\mathrm{s}=0$ the flow above the center of the intruder impinges on the intruder from the left side, while the flow below the intruder center impinges on the intruder from the right side, resulting in $F_\mathrm{L} = 0$. With increasing $u^*_\mathrm{s}$ in the granular case, figures~\ref{sigmaz}(a-f), the upward stress zone on the lower right part of the intruder shifts slightly to the right and increases in magnitude, while the downward stress zone on the upper part of the intruder first shifts from the left side of the intruder to the right side before increasing in magnitude. This is explained in terms of the position of the intruder relative to zero velocity in the shear velocity profile: with increasing $u^*_\mathrm{s}$, bed particles below the center of the intruder impinge on the intruder from the right at an increasing relative speed, while bed particles above the center of the intruder shift from impinging from the left for $0 \leq u^*_\mathrm{s} \leq 1$ to impinging from the right for $u^*_\mathrm{s} > 1$ (see figure~\ref{field}), with the number and relative speed of particles impacting from the right increasing with increasing $u^*_\mathrm{s}$. The shape of the contours in the granular case are more curved than those in the fluid case. In addition to the region of negative (downward) stress shifting from the trailing side to the leading side of the intruder with increasing $u^*_\mathrm{s}$, for $0.5\leq u^*_\mathrm{s} \leq 2$ the positive (upward) stress on the bottom of the intruder is stronger than the negative (downward) stress on the top of the intruder.  This  leads to $F_\mathrm{L} > 0$ in the granular case for $0 < u^*_\mathrm{s} < 3$, evident in figure~\ref{F_l_scaled_par_varied}. When $u^*_\mathrm{s}=5$, the downward stress is somewhat larger than the upward stress, resulting in $F_\mathrm{L}<0$.

In the fluid case, figures~\ref{sigmaz}(g-l), the vertical stress contours are nearly straight (circles in 3D) and aligned perpendicular to an axis tilted toward the left, independent of $u^*_\mathrm{s}$.  While the downward stress remains nearly constant with increasing $u^*_\mathrm{s}$, the upward stress weakens, causing $F_\mathrm{L}$ to become increasingly negative, which is consistent with the lift for the fluid case, shown in figure~\ref{F_LvsVsdimensional}.

\begin{figure}
\centerline{\includegraphics[width=\textwidth]{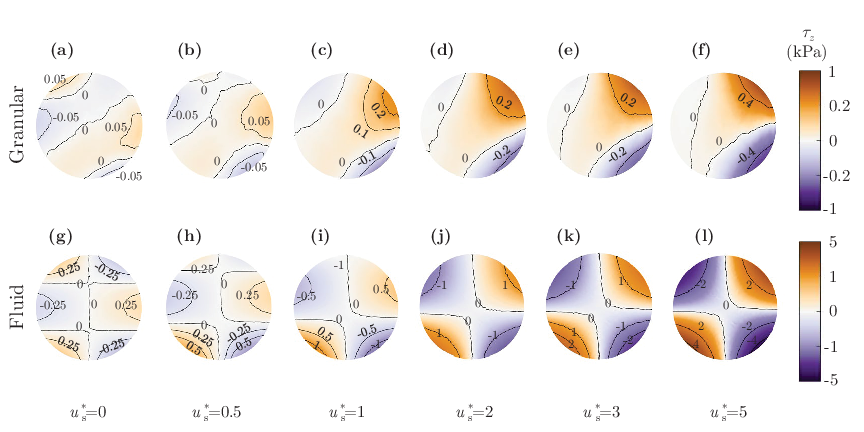}}
\caption{Distributions of intruder vertical shear stress component, $\tau_{z}$, in (a-f) granular flow and (g-l) fluid flow with increasing dimensionless slip velocity, $u^*_\mathrm{s}$. Note the different colorbar scales.}
  \label{sigmazt}
\end{figure}

To further understand the origin of the differences in $\sigma_z$ for the granular and fluid flow cases, we consider the distribution of vertical components of the shear stress, $\tau_{z}$, and the normal stress, $P_z$, separately. From figure~\ref{sigmazt} showing $\tau_{z}$ on the intruder surface, it is clear that the vertical shear stress in the granular case is as much as five times smaller than that for the fluid case. This likely occurs because the granular shear stress is bounded by $\mu_pP$ and results from intermittent sliding contact of bed particles on the intruder, while the viscous friction in a Newtonian fluid acts continually on the surface of the intruder. There is negligible shear on the trailing side of the intruder in the granular case because of the void that forms behind the intruder. Although the magnitudes are quite different, for $u^*_\mathrm{s}\geq 1$ the distributions of vertical shear on the leading side of the intruder share some similarities. In both cases, there is an upward shear component on the upper portion of the leading side as the flow slips upward over the intruder and a downward shear component on the lower portion of the leading side as the flow slips downward below the intruder.  

\begin{figure}
\centerline{\includegraphics[width=\textwidth]{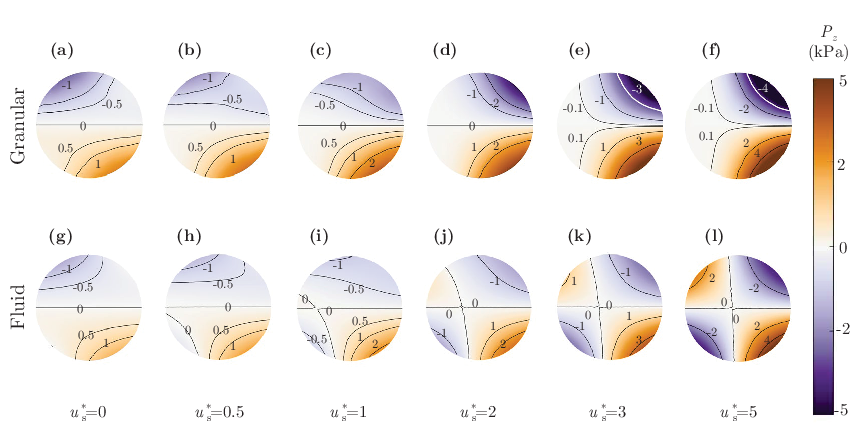}}
  \caption{Vertical component of normal stress on the intruder, $P_z$, in (a-f) granular flow and (g-l) fluid flow with increasing dimensionless slip velocity, $u^*_\mathrm{s}$.}
  \label{sigmazn}
\end{figure}

In contrast to $\tau_z$, the magnitude of $P_z$ is higher for the granular case than the fluid case, as shown in figure~\ref{sigmazn}. Specifically, the absolute value of $P_z$ averaged over the leading hemisphere for the granular case is 1.6 to 1.9 times higher than that for the fluid case, consistent with the pressure comparison in figure~\ref{stress_dist}. Since $\tau_z$ is so much smaller than $P_z$ for the granular case, $P_z$ nearly matches the distributions of the total stress, $\sigma_{z}$, for the granular case, in figures~\ref{sigmaz}(a-f) for all $u^*_\mathrm{s}$. Hence, the lift force in granular flow is primarily a consequence of the vertical component of the normal stress, with a considerably smaller contribution from $\tau_{z}$.  As described in the context of figure~\ref{sigmaz}, the distribution of $P_z$ is readily explained in terms of the position of the intruder relative to zero velocity in the shear velocity profile and the way that bed particles above and below the center of the intruder impinge on the intruder. $P_z$ in the fluid case, figures~\ref{sigmazn}(g-l), is similar to the granular case at $u^*_\mathrm{s}=0$, but is increasingly different as $u^*_\mathrm{s}$ is increased. $P_z$ on the leading (right) hemisphere of the intruder remains similar in the two cases with downward $P_z$ on the upper right and upward $P_z$ on the lower right. However, additional normal stress zones emerge on the trailing (left) hemisphere of the intruder in the fluid case for $u^*_\mathrm{s} \geq 0.5$, see figures~\ref{sigmazn}(h-l). In these zones, $P_z$ is opposite in sign to the granular cases and grows to be quite large in magnitude by $u^*_\mathrm{s} =5$. The signs of $P_z$ on the trailing side of the hemisphere in the fluid case are opposite to those on the leading side: $P_z$ on the top-left is upward and on the bottom-left is downward. 
In contrast, the pressure in the void behind the intruder in a granular flow cannot be negative.  Although the magnitude of $P_z$ on the bottom of the intruder, no matter upward or downward, is larger than the magnitude of $P_z$ on the top of the intruder in the fluid case, it is the combination of $P_z$ and $\tau_z$ that drives the lift force. In fact, $P_z$ and $\tau_z$ in the fluid case have opposite signs, similar magnitudes, and act on the entire intruder surface, as is evident by comparing figures~\ref{sigmazn}(g-l) to figures~\ref{sigmazt}(g-l). This is quite different from the granular case, where $\tau_z$ is an order of magnitude less than $P_z$ and both act primarily on the leading hemisphere of the intruder. 

\begin{figure}
\centerline{\includegraphics[width=\textwidth]{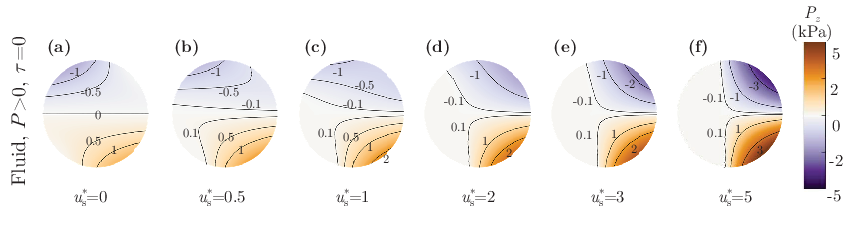}}
  \caption{For the $P>0$ pressure modification of fluid stresses with $\tau=0$ (see text), vertical component of total stresses on the intruder, $\sigma_z$, with increasing dimensionless slip velocity, $u^*_\mathrm{s}$.}
  \label{P_truncation}
\end{figure}

Given that $\tau_z$ is small compared to $P_z$ in the granular case and that the distributions of $P_z$ in the fluid and granular cases are similar on the leading hemisphere of the intruder, we return to the idea of artificially modifying the pressure and shear stress on the intruder in fluid shear flow to mimic that on an intruder in granular shear flow. Consider the distribution of the vertical components of stress, $\sigma_z$, for the fluid case only where $P >0$ with $\tau$ set to zero, shown in figure~\ref{P_truncation}. The similarity between this figure and the first row of figure~\ref{sigmaz} suggests that the differences in the lift force between the granular shear flow and the fluid shear flow are related to i) the relatively low shear stresses exerted by the flow on an intruder in granular shear flow, and ii) the ability of the fluid to exert a negative gauge pressure at positions on the intruder where there is a void in a granular shear flow.

We return now to figure~\ref{modification1}(c) for the lift force to describe how including only $P>0$ regions and setting $\tau = 0$ for the fluid case compares to the granular case.  Like the simpler `leading hemisphere' approach, $F_\mathrm{L}$ changes from negative to positive, matching the sign for $F_\mathrm{L}$ in granular shear flow for small $u_\mathrm{s}$\,m/s. 
Unlike the `leading hemisphere' approach, the $(P>0,\tau=0)$-modification correctly gives $F_\mathrm{L}=0$ at $u_\mathrm{s}=0$.  However, as $u_\mathrm{s}$ increases, the approach asymptotes to a value that is about one-half
the peak value in the granular case and fails to become negative for large $u_\mathrm{s}$, as it does in the granular case.


We have implemented more complicated modifications to the fluid shear flow stresses to better match the granular shear flow lift force including using shear stresses based on the $\mu(I)$-rheology, accounting for the bed particle diameter by considering the pressure and shear on a spherical shell a distance $d/2$ beyond the intruder surface, and combinations of these approaches. Although $F_\mathrm{L}$ at $u^*_\mathrm{s}=1$ for some approaches reaches a magnitude closer to the peak value of $F_\mathrm{L}$ in the granular case, none are significantly more successful than the two simpler approaches described above because $F_\mathrm{L}$ in the modified cases remains positive as $u^*_\mathrm{s}$ becomes large. This failure to reproduce the granular shear flow lift force by modifying the fluid results demonstrates the sensitivity of $F_\mathrm{L}$ to subtle differences in the interaction of the flow field with the intruder. Despite the general similarity between the modified stress distribution for the fluid case in figure~\ref{P_truncation} and the stress distribution for the granular case in figures~\ref{sigmaz}(a-f), details of the stress distributions significantly impact the lift force, $F_\mathrm{L}$. 

\subsection{Effects of rotation on lift}
\begin{figure}
  \centerline{\includegraphics{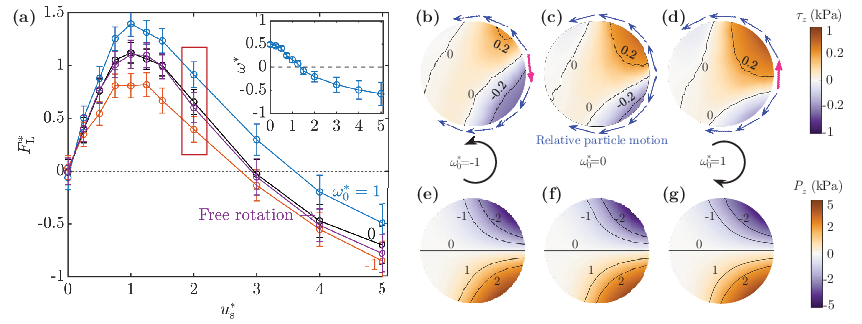}}
  \caption{(a) Dimensionless lift force, $F^{*}_\mathrm{L},$ vs.\ dimensionless slip velocity, $u^*_\mathrm{s},$ for dimensionless angular velocities, $\omega^*_0=\pm 1$, and a freely rotating intruder ($R=3$, $P_0= 1000$\,Pa, $\dot\gamma_0=5$\,s$^{-1}$). Inset: scaled angular velocity $\omega^*$ vs.\ $u^*_\mathrm{s}$ for freely rotating intruder.  Vertical component of (b-d) shear stress, $\tau_z$, and (e-g) normal stress on the intruder, $P_z$, in granular flow with no rotation and with imposed positive and negative rotations at $u^*_\mathrm{s}=2$ (red box in (a)). Fields in (c) and (f) are the same as figure~\ref{sigmazt}(d) and figure~\ref{sigmazn}(d), respectively. The vectors (not to scale) compare relative slip direction of contacting bed particles about an intruder for different $\omega^*_0$.  Red vectors indicate where imposed rotation reverses slip velocity direction, and, consequently, the tangential stress direction.}
  \label{Rotation}
\end{figure}

Up to this point, the intruder has been constrained to not rotate. For completeness, here we impose angular velocity, $\omega_{0}$, on the intruder about the $y$-axis, as well as allowing the intruder to freely rotate with angular velocity $\omega$ about the $y$-axis in response to the granular shear flow. Figure~\ref{Rotation}(a) shows the dependence of $F^*_\mathrm{L}$ on $u^*_\mathrm{s}$ for different values of the imposed dimensionless angular velocity $\omega_0^*=\omega_0/\dot\gamma_0$. Other parameters are the same as in Sec.~\ref{Typical_case}.  For $\omega^*_0=1$ (clockwise rotation), $F^*_\mathrm{L}$ shifts upward with respect to the non-rotating case, and when $\omega^*_0=-1$ (counter-clockwise rotation), $F^*_\mathrm{L}$ shifts downward.  Near the peak in $F^*_\mathrm{L}$, the $u^*_\mathrm{s}$ dependent shifts are nearly symmetric with respect to the non-rotating case, while for $u^*_\mathrm{s}\geq 3$, the upward shift for $\omega^*_0=1$ is 2-3 times larger than the downward shift for $\omega_0=-1$. For a freely rotating intruder, $F^*_\mathrm{L}$ nearly matches the case where the intruder is not allowed to rotate, except, perhaps, at $u^*_\mathrm{s}=5$, where $F^*_\mathrm{L}$ is slightly lower.

From the comparison of shear stress, $\tau_z$, shown in figures~\ref{Rotation}(b-d), and normal stress on the intruder, $P_z$, in figures~\ref{Rotation}(e-g), for different $\omega^*_0$ at $u^*_\mathrm{s}=2$, it is apparent that the changes in the lift force with intruder rotation are due to changes in shear stress as the intruder slips relative to the surrounding bed particles, while the normal stress is apparently unaffected by intruder rotation. The vectors on the intruder periphery in figures~\ref{Rotation}(b-d) illustrate how the slip direction changes with intruder rotation.  At $\omega^*_0=0$ bed particles flowing over the intruder separate to flow above or below the intruder at approximately the middle of the intruder. When the intruder is rotated clockwise ($\omega^*_0=1$), the zero slip separation point shifts downward (red vector indicates change in slip direction), increasing the upward vertical frictional force and shifting $F^*_\mathrm{L}$ higher. The opposite occurs for $\omega^*_0=-1$, where the separation point shifts upward, increasing the total downward vertical frictional force and shifting $F^*_\mathrm{L}$ lower. 
 
Returning to the free rotation case, the inset in figure~\ref{Rotation}(a) shows the dimensionless mean angular velocity $\omega^*=\omega/\dot\gamma_0$ for a freely rotating intruder. For $u^*_\mathrm{s}=0$, the intruder rotates with a mean angular velocity of $\dot\gamma_0/2$ corresponding to $\omega^*=0.5$, which matches previous results in granular \citep{da2005rheophysics} and laminar fluid shear flows \citep{kohlman1965measurement}, and is due to the freely rotating intruder's lack of resistance to the rotational component of the shear flow \citep{guazzelli2018rheology}. For $u^*_\mathrm{s}>0$, the intruder rotates with the unperturbed shear (clockwise) for $u^*_\mathrm{s}<1$ and against the shear (counterclockwise) for $u^*_\mathrm{s}>1$.



\section{Conclusions}\label{sec:conclusions}
Frequently, details of granular and fluid flows for the same geometry and forcing vary due to their differing rheologies, but their qualitative flow features are often similar.  In particular, the lift force on a spherical particle in a granular shear flow would be expected to be similar in nature to the Saffman lift force in a fluid shear flow based on other similarities between granular and fluid flows, such as the Stokes-like drag force (figure~\ref{Drag_coeff}).  It is therefore surprising that the sign of the granular lift force is opposite to the corresponding fluid lift force for small slip velocities (figure~\ref{F_LvsVsdimensional}), which leads an untethered intruder particle to move to where the slip velocity is zero (figure~\ref{z-track}).
Our results showing the lift force to be a non-monotonic function of the slip velocity are robust and can be collapsed to a single master curve by non-dimensionalizing both quantities with respect to the particle sizes, shear rate, and overburden pressure (figure~\ref{F_l_scaled_par_varied}). Similarly, the drag force (figure~\ref{Drag_coeff}) data can be collapsed and match well with previous Stokes-like drag results for an intruder crossing flow streamlines even though here the intruder moves parallel to the streamlines. Analogous fluid simulations show that the flow field around an intruder is very similar to that for the corresponding granular shear flow (figure~\ref{field}).

The difference in the lift force between a granular shear flow and a fluid shear flow lies in the details of the flow of particles or fluid near and in direct contact with the intruder particle, most plainly evident in the region void of particles in the granular case (figure~\ref{field}) that results in distinct pressure fields on the surface of the intruder particle for the granular and fluid cases (figure~\ref{stress_dist}). However, it is not just the negative pressure on the trailing side of the intruder in the fluid case that differs from the granular case. Differences are also related to the much weaker shear stress component in the vertical direction in the granular case (figure~\ref{sigmazt}) relative to the fluid case. The stronger shear stress in the fluid case offsets the pressure in the fluid case (figures~\ref{sigmazt} and \ref{sigmazn}) resulting in different net vertical stresses between the two cases (figure~\ref{sigmaz}). Although we modify the fluid solution to mimic the granular lift force results by accounting for the void behind the intruder and the much smaller shear stresses in the granular case, this is only partially successful in that it demonstrates the reversal of the sign of the lift force at low slip velocities but does not match the sign of the lift force at higher slip velocities. Clearly, differences in the fluid and granular rheologies strongly influence the interactions with an intruder in shear flows. While the differences are minimal for the drag force, which is large compared to the lift force (figure~\ref{Lift_drag_ratio}), subtle variations in the vertical stresses acting on the intruder have a significant impact on the lift force.

While we have made progress in understanding the lift force on an intruder in a granular shear flow, much remains to be learned. Computational constraints make it difficult to pursue simulations at large size ratios and large slip velocities, both of which require a large computational domain. Yet both of these regimes would be helpful in bringing the granular simulations closer to the limiting case of a fluid flow. It would also be worthwhile to pursue analogous simulations of the effectively two-dimensional case of a cylinder in a granular shear flow. Preliminary simulations indicate that the lift force on a cylinder is similar to that for a spherical intruder in that it has a similar sign and magnitude, although the lift force remains positive for much larger slip velocities than for a cylinder. Finally, experiments to verify these lift force results would be invaluable, although this would likely be difficult given the usual challenges in measuring forces on individual particles in granular flows as well as the small magnitude of the lift force and the stochastic nature of granular flows. Although questions remain, progress is being made in the understanding of buoyancy, drag, and, now, lift forces on intruders in granular flows.


\section*{Acknowledgements}
\label{sec:acknowledgemets}
We thank Lu Jing, Yifei Duan, and Gregory J.\ Wagner for valuable discussions. The research was facilitated in part by the computational resources and professional services provided by the Quest high performance computing facility at Northwestern University, which is jointly supported by the Office of the Provost, the Office for Research, and Northwestern University Information Technology.

\section*{Funding}
\label{sec:funding}
This material is based upon work supported by the National Science Foundation under Grant No.\ CBET-1929265, a Grant Opportunities for Academic Liaison with Industry (GOALI) program in collaboration with The Dow Chemical Company and The Procter \& Gamble Company. 

\section*{Declaration of Interest}
\label{sec:declaration_interest}
The authors report no conflict of interest.

\appendix
\section{Equivalence of Lift Force Characterization Methods}\label{app_comparison_F_L_methods}

\begin{figure}
\centerline{\includegraphics{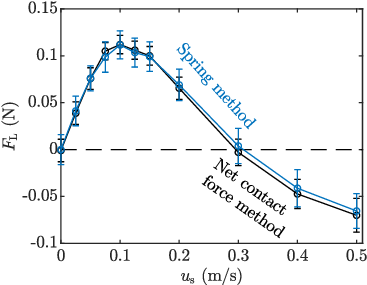}}
  \caption{Comparison of lift force, $F_\mathrm{L}$, vs.\ slip velocity, $u_\mathrm{s}$, measured using the virtual spring method~\citep{guillard2016scaling} (blue) and the net vertical force due to bed particle contacts (black) for $R=3$, $P_0=1$\,kPa, and $\dot\gamma_0=5$\,s$^{-1}$.}
  \label{Methods_Comparison}
\end{figure}

As described in Sec.~\ref{sec:methods}, the lift force, $F_\mathrm{L}$, is calculated as the net vertical force on the intruder due to contacts with bed particles, typically averaged over 4\,s. An alternate method to measure the lift force tethers the intruder to a virtual spring~\citep{guillard2016scaling} that acts only in the $z$-direction. We measure the lift force on the intruder with this method for a typical case with $R=3$, $P_0=1$\,kPa, and $\dot\gamma_0=5$\,s$^{-1}$.  The virtual spring stiffness, $k=200$\,N/m constrains the vertical displacement to less than $d/10$. The average vertical displacement of the intruder, $z_{\mathrm{i, s}}$, is measured over 4\,s, and determines the spring-based lift force, $F_{\mathrm{L, s}}=kz_{\mathrm{i, s}}$.  Figure~\ref{Methods_Comparison} compares the lift force results using both methods and indicates that no significant differences exist across the range of $u_\mathrm{s}$ considered here.

\section{Previous measurements of granular lift}\label{Saffman}

\begin{figure}  
\centerline{\includegraphics{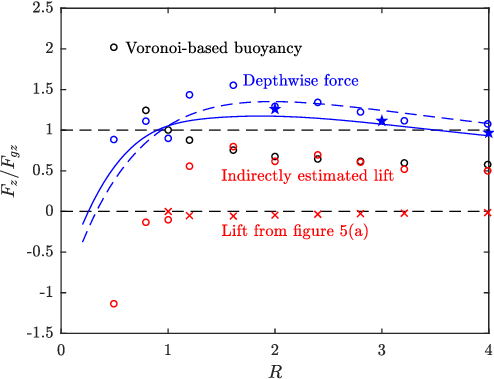}}
\caption{Comparison of segregation force $F_z$ normalized by intruder weight in depthwise-direction, $F_{gz}$, vs.\ the size ratio, $R$. Data from \citet{van2018segregation} ($\circ$) show the total depthwise force (blue), the Voronoi-based buoyancy force (black), and the indirectly estimated lift force due to slip velocity (red) and are compared to the lift force calculated using figure~\ref{F_l_scaled_par_varied}(a) and the depthwise total force ($\star$) in \citet{yennemadi2023drag}. The blue solid curve shows the predicted total segregation force and the blue dashed curve shows the buoyancy due to the pressure gradient based on \citet{jing2021unified}.}
  \label{compare_studies}
\end{figure}

In a study focused on segregation forces in gravity-driven chute flows, \citet{van2018segregation} indirectly estimated the lift force at small slip velocities, $\left|u^*_\mathrm{s}\right|\leq 0.01$. Their results in figure~\ref{compare_studies} for a range of $R$ show the indirectly estimated lift force (red $\circ$) determined as the difference between the measured depthwise total force (blue $\circ$) and an estimated buoyancy force (black $\circ$), which is based on the free volume around the intruder. For comparison, the lift force calculated using the scaled lift force curve in figure~\ref{F_l_scaled_par_varied}(a) is also shown in figure~\ref{compare_studies} (red $\times$) for $1 \leq R\leq 4$ based on an estimate of the streamwise slip velocity (equation (2) in \citet{van2018segregation}). The lift force from figure~\ref{F_l_scaled_par_varied}(a) is opposite in sign to that found in \citet{van2018segregation} with a magnitude at least ten times smaller. We also include in figure~\ref{compare_studies} the segregation force (blue solid curve) proposed by \citet{jing2021unified}, which is the sum of the buoyancy force due to the pressure gradient (blue dashed curve) and the segregation force due to the shear rate gradient (difference between the dashed and solid blue curves). It is evident that the depthwise total force measured by \citet{van2018segregation} is quite similar to the segregation force model (blue solid curve), which is mostly due to the buoyancy force (blue dashed curve). Furthermore, the depthwise total force measurements (blue $\star$) in gravity-driven flow in \citet{yennemadi2023drag} are also consistent with the segregation force model \citep{jing2021unified}. Hence, it appears that the large Saffman-like lift force found by \citet{van2018segregation} is a consequence of the approach used to estimate the buoyancy force.

\section{Lift forces in fluids}\label{Fluid_lift}

The factors contributing to positive and negative lift forces on intruders in fluid shear flow fail to explain the lift force in granular shear flow.  However, it is helpful to consider these factors because the lift force on an intruder in unbounded fluid shear flow has been studied in detail~\citep{saffman1965lift, mclaughlin1991inertial, magnaudet1998some, stone2000philip, shi2019lift}.
The lift force on an intruder in a sheared fluid originates in inertial effects even at low Reynolds numbers; no lift force will be predicted without considering inertial effects~\citep{bretherton1962motion}.  Viscous and inertial effects due to shear and slip motion decay differently with distance, and their competition creates two independent length scales: $l_\mathrm{s}=\frac{\eta}{\rho_f u_\mathrm{s}}$ and $l_{\gamma}=\sqrt{\frac{\eta}{\rho_f \dot\gamma_0}}$.
Viscous effects are dominated by slip beyond $l_\mathrm{s}$ and by shear beyond $l_{\gamma}$. Corresponding to the two length scales, two independent Reynolds numbers can be defined as $Re_\mathrm{s}=\frac{\rho_fu_\mathrm{s}r_\mathrm{i}}{\eta}=\frac{r_\mathrm{i}}{l_s}$ and $Re_\gamma=\frac{\rho_f\dot{\gamma}_0r_\mathrm{i}^2}{\eta}=\frac{r_\mathrm{i}^2}{l_\gamma^2}$, where $r_\mathrm{i}=d_\mathrm{i}/2$ is the intruder radius, with the ratio $\epsilon=\sqrt{Re_\gamma}/Re_\mathrm{s}=l_\mathrm{s}/l_\gamma$.  
In his classic analysis, \citet{saffman1965lift} assumed $r_\mathrm{i}\ll l_\gamma\ll l_\mathrm{s}$ (therefore, $\epsilon\gg 1$), or equivalently $Re_\mathrm{s}\ll \sqrt{Re_\gamma}\ll1$, so that the slip-related inertial effect can be neglected. With the flow field divided into an inner region dominated by viscous effects and an outer region dominated by inertial effects, a matched asymptotic expansion analysis gives the lift force as 
\begin{eqnarray}
\frac{\rho_fF_\mathrm{L}}{\eta^2}=-\frac{9}{\pi}J_\infty\sqrt{Re_\gamma}Re_\mathrm{s}+\frac{11\pi}{8}Re_\gamma Re_\mathrm{s},
\label{saffman}
\end{eqnarray}
where $J_\infty=2.255$. On the r.h.s., the first term dominates since $\sqrt{Re_\gamma}\ll1$.  If the intruder particle is allowed to rotate, the intruder rotational velocity, $\omega,$ is of the order of $\dot{\gamma}_0$ ($\omega\approx\dot{\gamma}_0/2$ for a free spinning intruder). The additional lift from rotation is $-\pi\rho_f \omega u_\mathrm{s} r_\mathrm{i}^3$, which is consistent with the lift force on an intruder in uniform fluid flow \citep{rubinow1961transverse} but is an order of magnitude smaller than the lift in (\ref{saffman}).
As a result of the small effects of the second term in (\ref{saffman}) and the effect of particle rotation, many studies consider only the first term on the r.h.s.\ of (\ref{saffman}).
With the constraints $Re_\mathrm{s}\ll \sqrt{Re_\gamma}\ll1$ relaxed, a recent review \citep{shi2019lift} proposed a more generalized lift force correlation based on data in the literature over a broader range ($Re_\mathrm{s}<50$):
\begin{eqnarray}
\frac{\rho_fF_\mathrm{L}}{\eta^2}=-\frac{9}{\pi}J(\epsilon)\sqrt{Re_\gamma}Re_\mathrm{s}
+\frac{11\pi}{8}Re_\gamma Re_\mathrm{s} \exp(-Re_\mathrm{s}),
\label{Saffman_improved}
\end{eqnarray}
where $J(\epsilon)$ is a fit to the numerical evaluation of the integral provided by \citet{mclaughlin1991inertial}, whose asymptotic expansions included the inertial effect on the slip motion (therefore relaxing the $\epsilon=\sqrt{Re_\gamma}/Re_\mathrm{s} \gg 1$ constraint in Saffman's analysis). 
$J(\epsilon)$ recovers $J(\infty)=J_\infty=2.255$ and is negative for $\epsilon<0.22$:
\begin{equation}
 J(\epsilon) =
    \begin{cases}
    -0.04\epsilon+2.05\epsilon^2-32.2\epsilon^3+106.8\epsilon^4 & \text{for $\epsilon\leq 0.23$,}\\    2.255(1+0.02304\epsilon^{-2})^{-12.77} & \text{for $\epsilon> 0.23$.}
    \end{cases}      
    \label{J_formula}
\end{equation}
However, the validity of McLaughlin's asymptotic expansions for $\epsilon<0.7$ has been questioned~\citep{legendre1998lift}, and, as concluded in the review by \citet{shi2019lift}, $J(\epsilon)<0$ is never observed in numerical simulations or experiments (including the experimental work co-authored by Mclaughlin~\citep{cherukat1994inertial}) at low $Re_\mathrm{s}$, leading to $F_\mathrm{L}<0$ as shown in figure~\ref{F_LvsVsdimensional}. The second term on the r.h.s.\ of (\ref{Saffman_improved}) accounts for large values of $Re_\mathrm{s}$ via the exponential decay term. In fact, only for $Re_\mathrm{s}\geq60$, did \citet{kurose1999drag} find that $F_\mathrm{L}$ changes sign to become positive.

For our fluid simulations with $P_0=\SI{1000}{Pa}$, $\dot{\gamma}_0=\SI{5}{s^{-1}}$ and $R=3$, the slip velocity $u_\mathrm{s}\leq\SI{0.5}{m/s}$ or $u^*_\mathrm{s}\leq5$, giving $Re_\mathrm{s}\leq0.07$, $Re_{\gamma}=0.0052$ and $\epsilon\geq1.04$. In this range of dimensionless parameters, the fluid lift force should be well predicted by the first term on the r.h.s.\ of (\ref{Saffman_improved}) with $J(\epsilon)$ from (\ref{J_formula}), which is shown as the solid red curves in figures~\ref{F_LvsVsdimensional} and \ref{modification1}.

In addition to fluids, lift forces have also been studied in rarefied gases where the mean free path $\lambda$ is larger than the intruder size as described by the Knudsen number ${K_n}=\lambda/r_\mathrm{i}\gg1$.  \citet{kroger2006unifying} find the sign of the lift force to be positive, opposite to that in continuous fluid \citep{saffman1965lift} but matching our results for granular shear flow at low slip velocities. \citet{liu2008forces, liu2009forces} treat the collisions between gas molecules and the intruder particle as rigid-body collisions and give analytical results for the positive lift force with a magnitude that depends on how diffuse or specular the collisions are. Considering non-rigid-body collisions, \citet{luo2016lift} find that the gas temperature influences the scattering angle and the lift force reverses as the temperature increases even at a fixed shear rate and slip velocity. Similarly in granular shear flow, the scattering angle may vary with slip velocity or granular temperature, which further influences the efficiency of momentum exchange on the top and bottom hemispheres of the intruder, and therefore affects the lift force. Knowing that the lift forces have different signs in the continuum fluid and rarified gas limits, one question is when does the transition from negative to positive $F_{\mathrm{L}}$ happen. A recent study by \citet{taguchi2022inversion} using a simpler setup with a rotating intruder slipping in a uniform flow (no shear) measured the lift forces at various Knudsen numbers and found that the transition occurs at $K_n=0.71$. However, it remains unclear whether or not such a transition exists in shear flows.

Returning to (\ref{saffman}), one might speculate that the scaling relation in the equation applies to lift on an intruder for granular shear flow. As we show in figure~\ref{F_l_scaled_par_varied}, even though the DEM simulations cover a range of values for $I$ and $R$, the data still collapse onto a master curve when plotted in terms of $F^*_\mathrm{L}=F_\mathrm{L}/\left(P_0(d_\mathrm{i}+d)^2/4\right)$ and $u^*_\mathrm{s}=u_\mathrm{s}/(\dot\gamma_0(d_\mathrm{i}+d))$ in granular shear flow. 
This raises the question of whether the scaling relation for the lift force in fluid shear flow in (\ref{saffman}) can be connected to the scaling relation in granular shear flow. Here, considering only the leading term (first term on the r.h.s.) in (\ref{saffman}), we follow the same non-dimensionalization with $P_0(d_\mathrm{i}+d)^2/4$:
\begin{eqnarray}
\frac{F_\mathrm{L}}{P_0 (d_\mathrm{i}+d)^2/4}=-\frac{9}{\pi} J_\infty \frac{\sqrt{\rho_f\dot{\gamma_0}\eta}}{P_0}
\frac{R^2}{(R+1)^2}
u_\mathrm{s}.
\label{saffman_scaling1}
\end{eqnarray}
Substituting $\rho_f=\rho\phi$, $\eta=\mu(I)P_0/\dot{\gamma_0}$, and $I=\dot\gamma_0d/\sqrt{P_0/\rho}$ gives
\begin{eqnarray}
F_\mathrm{L}^*=-\frac{9}{\pi} 
J_\infty
\sqrt{\mu\phi}I\frac{R^2}{R+1}
u_\mathrm{s}^*.
\label{saffman_scaling2}
\end{eqnarray}
(\ref{saffman_scaling2}) clearly demonstrates the dependence of the fluid lift force on the inertial number $I$ and size ratio $R$ for a corresponding granular flow, even if we neglect the variance on $\mu$ and $\phi$. Thus, the scaling relation in granular shear flow seems unrelated to that for fluid shear flow. That the scaling in (\ref{saffman}) developed for lift in a fluid shear flow cannot does not apply to the granular shear flow case is not surprising given the differences in the vertical components of shear and normal stresses evident in figures~\ref{sigmazt} and~\ref{sigmazn}.

\bibliographystyle{jfm}
\bibliography{jfm-instructions}

\end{document}